\documentclass[aps,pre,notitlepage,showpacs]{revtex4-1}

\usepackage{graphicx}
\usepackage{amsmath}

\newcommand{\cH}{{\cal H}}
\newcommand{\cK}{{\cal K}}
\newcommand{\cL}{{\cal L}}
\newcommand{\cS}{{\cal S}}

\newcommand{\eg} {{e.g., }}

\newcommand{\half} {\frac{1}{2}}
\newcommand{\ie} {{i.e., }}
\newcommand{\Mf}{M_{\rm f}}
\newcommand{\pd} {\partial}
\newcommand{\tu}{\tilde{u}}

\begin{document}

\title{Shape and symmetry of a fluid-supported elastic sheet}

\author{Haim Diamant} 
\email{hdiamant@tau.ac.il} 
\affiliation{Raymond \& Beverly Sackler School of Chemistry, Tel Aviv
University, Tel Aviv 69978, Israel}

\author{Thomas A.\ Witten} 
\email{t-witten@uchicago.edu}
\affiliation{Department of Physics and James Franck Institute,
University of Chicago, Chicago, Illinois 60637, USA}

\date{July 7, 2013}

\begin{abstract} 
A connection between the dynamics of a sine-Gordon chain and a certain
static membrane folding problem was recently found.  The
one-dimensional membrane profile is a cross-section of the
position-time sine-Gordon amplitude profile.  Here we show that when
one system is embedded in a higher-dimensional system in this way,
obvious symmetries in the larger system can lead to nontrivial
symmetries in the embedded system.  In particular, a thin buckled
membrane on a fluid substrate has a continuous degeneracy that
interpolates between a symmetric and an antisymmetric fold.  We find
the Hamiltonian generator of this symmetry and the corresponding
conserved momentum by interpreting the simple translational symmetries
of the sine-Gordon chain in terms of the embedded coordinates.  We
discuss possible extensions to other embedded dynamical systems.
\end{abstract}


\maketitle

\section{Introduction}
\label{sec_intro}

Thin elastic sheets form rich patterns when stressed or confined. One
may try to account for these patterns as weak deformations of a simple
(\eg flat) reference state. It has been recognized, however, that as
the sheet is made increasingly thin and bendable, the validity range
of such near-threshold analyses shrinks indefinitely, along with their
experimental relevance
\cite{Davidovitch11,Davidovitch12,King12}. Crumpled paper is a prime
example of such a far-from-threshold behavior \cite{Witten07}.

A thin elastic sheet, compressed on top of a heavy fluid
\cite{Milner89,Cerda03,Vella04,Huang07,Pocivavsek08,Leahy10,Huang10,Holmes10,arxiv10,Audoly11,prl11,sm13},
provides a simpler example of this situation. The leading-order
analysis of the deformation yields a periodic pattern of wrinkles with
wavelength $\lambda=2\pi[B/(\rho g)]^{1/4}$
\cite{Milner89,Cerda03}. The finite wavelength arises from a
competition between the resistance of the sheet to bending ($B$ being
the sheet's bending rigidity) and the resistance of the fluid
substrate to height changes ($\rho$ being the fluid mass density and
$g$ the gravitational acceleration). Since $B$ depends on the sheet
thickness $H$ as $B\sim H^3$, the wrinkling wavelength decreases with
decreasing thickness as $\lambda\sim H^{3/4}$. A higher-order analysis
reveals that the wrinkling pattern is actually stable only for small
lateral displacements, $\Delta<\Delta_w\sim\lambda^2/L$, where
$L$ is the total length of the sheet \cite{arxiv10,Audoly11}. This
validity range diminishes with either decreasing $H$ or increasing
$L$. Beyond it the deformation becomes localized in a finite domain
(fold) of width $\kappa^{-1}\sim\lambda^2/\Delta$
\cite{arxiv10,Audoly11}. It takes a mere displacement of
$\Delta\sim\lambda$ to concentrate all the deformation into a single
wavelength and make the fold contact itself
\cite{Pocivavsek08,arxiv10,prl11,sm13}. Thus, the relevant response of a
sufficiently thin sheet is practically always far from threshold,
involving a strong localized deformation. Similar observations hold
when the fluid substrate is replaced by a linear elastic one
\cite{Audoly11,Thompson,Hunt93,LeeWaas96,Zhang07,Audoly08,Reis09,Brau11}.

It is therefore fortunate and beneficial that the fluid-supported
sheet under uniaxial compression turns out to be an integrable system
\cite{prl11}. Exactly solvable problems often owe their integrability
to extra symmetries. In the current work we focus on the corresponding
extra symmetry in the fluid-supported sheet, which is manifest in a
continuous degeneracy of ground states \cite{aps2012,Rivetti2013}. We
connect this peculiar property of the one-dimensional buckling problem
to the simple translational symmetries of the two-dimensional
sine-Gordon problem.

In Secs.\ \ref{sec_system} and \ref{sec_dynamical} we repeat the
definition of the problem and its representation as a dynamical system
\cite{prl11}. In Sec.\ \ref{sec_sG} we establish the relation between
our problem and the sine-Gordon chain. In Sec.\ \ref{sec_K} we
characterize the resulting extra symmetry, including consequences
concerning the fluid mass displaced by the sheet and the associated
buoyant force. Finally, in Sec.\ \ref{sec_discuss} we discuss
consequences of the new symmetry and analogous cases in other physical
systems.

\section{The system}
\label{sec_system}

\begin{figure}[tbh]
\vspace{0.3cm}
\centerline{\resizebox{0.5\textwidth}{!}
{\includegraphics{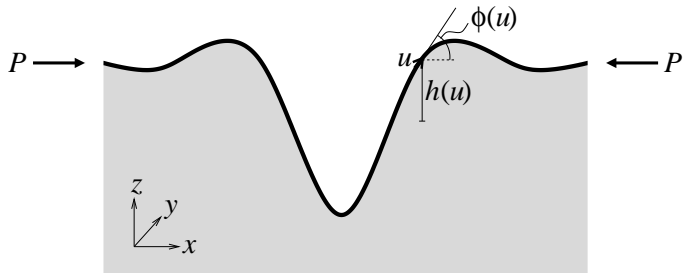}}}
\caption[]{Schematic view of the system.}
\label{fig_system}
\end{figure}

The system under consideration is schematically illustrated in Fig.\
\ref{fig_system}. An incompressible sheet of bending modulus $B$ lies
on a semi-infinite fluid of mass density $\rho$. The sheet is assumed
to be indefinitely long and deform in the $xz$ plane, remaining
uncurved along the $y$ axis. The deformation is parameterized by
either the height profile $h(u)$ or the slope angular profile
$\phi(u)$ as a function of the arclength $u$ along the sheet. The two
functions are geometrically linked by the relation $\dot h=\sin\phi$,
where hereafter a dot denotes a derivative with respect to $u$. The
sheet is subjected at its edges to a uniaxial pressure $P$ along $x$,
which is accompanied by a displacement $\Delta=\int_{-\infty}^\infty
du(1-\cos\phi)$ between the two edges.

A given deformation costs energy per unit length, $E=E_{\rm B}+E_{\rm
  G}$, with contributions from bending, $E_{\rm
  B}=(B/2)\int_{-\infty}^\infty du \dot\phi^2$, and from gravity,
$E_{\rm G}=(K/2)\int_{-\infty}^\infty du h^2\cos\phi$, where $K=\rho
g$. We use dimensionless variables, where all lengths are scaled by
$(B/K)^{1/4}=\lambda/(2\pi)$ and all energies by $B$. The pressure is
consequently scaled by $(BK)^{1/2}$. An equilibrium shape of the sheet
is one that minimizes $E-P\Delta$ for a given $P$ and the appropriate
boundary conditions. We thus look for the minimum of the following
functional:
\begin{align}
  \cS &= \int_{-\infty}^\infty du \cL(\phi,h,\dot\phi,\dot h)
  \notag\\ \cL &= \half\dot\phi^2 + \half h^2\cos\phi -
  P(1-\cos\phi) - Q(\sin\phi-\dot h).
\label{action}
\end{align}
In Eq.\ (\ref{action}) a Lagrange multiplier, $Q(u)$, has replaced the
local constraint on the relation between $h$ and $\phi$.

\section{The sheet as a dynamical system}
\label{sec_dynamical}

We study the deformation of the sheet using a mechanical analogy,
where the arclength $u$ along the sheet stands for time, and the angle
$\phi(u)$ and height $h(u)$ represent two coordinates with conjugate
momenta $p_\phi$ and $p_h$ \cite{prl11}. The equilibrium shape of the
sheet is given by the trajectory $[\phi(u),h(u)]$ that minimizes the
action in Eq.\ (\ref{action}). The Lagrangian
$\cL(\phi,h,\dot\phi,\dot{h})$ appearing in Eq.\ (\ref{action}) 
transforms into the following Hamiltonian:
\begin{equation}
  \cH(\phi,h,p_\phi,p_h) = 
  \half p_\phi^2 + p_h\sin\phi - \half h^2\cos\phi + P(1-\cos\phi).
\label{H}
\end{equation}
The fact that $\cH$ is a constant of motion corresponds to one
apparent symmetry of the homogeneous sheet\,---\,its invariance to
translation in $u$. In what follows we explicitly distinguish between
derivatives with respect to the canonical variables (using the symbol
$\pd$) and derivatives with respect to the coordinates (using dots and
primes).

Hamilton's equations yield the following dynamical system:
\begin{subequations}
\label{dynamical}
\begin{align}
  \dot\phi &= \pd_{p_\phi} \cH = p_\phi \label{dyn1}\\
  \dot h &= \pd_{p_h} \cH = \sin\phi \label{dyn2}\\
  \dot p_\phi &= -\pd_\phi \cH = -p_h\cos\phi - (h^2/2+P)\sin\phi \label{dyn3}\\
  \dot p_h &= -\pd_h \cH = h\cos\phi, \label{dyn4}
\end{align}
\end{subequations}
whose solution gives the equilibrium shape. Equations (\ref{dyn1}) and
(\ref{dyn4}) identify the $\phi$-conjugate momentum as the local
curvature, $p_\phi=\dot\phi$, and the $h$-conjugate momentum as the
fluid mass displaced by the deformed sheet up to point $u$,
$p_h=\int_{-\infty}^u du_1 h\cos\phi$. In the case of a localized fold
in an infinite sheet the following boundary conditions apply:
$h,\phi,\dot\phi=p_\phi=0$ at $u\rightarrow\pm\infty$, resulting in
$\cH=0$. Further requiring that $\ddot\phi=0$ far away from the fold
implies, through Eq.\ (\ref{dyn3}), that also $p_h=0$ at
$u\rightarrow\pm\infty$.  We have found the exact solution of this
problem by relating it to the integrable sine-Gordon chain
\cite{prl11}. The integrability of our system implies the existence of
a complete set of conserved canonical momenta. Since the system has
two canonical momenta, it must thus have two conserved dynamical
variables, the Hamiltonian $\cH$ and one other.  We shall call this
conserved quantity $\cK$ and determine it in Sec.\ \ref{sec_K}. We
need to discuss first the relation between our system and the
sine-Gordon chain.

\section{Relation to the sine-Gordon chain}
\label{sec_sG}

Equations (\ref{dyn1}), (\ref{dyn3}), and (\ref{H}) with $\cH=0$, yield
\begin{equation}
  \dddot\phi + (\dot\phi^2/2 + P)\dot\phi + h = 0,
\label{Euler}
\end{equation}
which is the known equation of Euler's {\it elastica}
\cite{Cerda05,Witten07}, with the hydrostatic pressure term $h$ ($\rho
g h$ in dimensional terms) playing here the role of an external normal
force.  Differentiating once with respect to $u$, we obtain a
fourth-order equation for the angle $\phi(u)$ alone,
\begin{equation}
  \ddddot\phi + [(3/2)\dot\phi^2 + P]\ddot\phi + \sin\phi = 0.
\label{fold}
\end{equation}

Equation (\ref{fold}) is related to a known hierarchy of nonlinear
partial differential equations\,---\,the combined sine-Gordon and
modified KdV (sG-mKdV) hierarchy \cite{Gesztesy}. The first three
equations in that hierarchy, for the two-dimensional function
$\phi(u,v)$, read:
\begin{subequations}
\begin{align}
  &\dot\phi' + 2(\beta e^{i\phi} - \alpha e^{-i\phi}) = 0 
 \label{sGmKdV1}\\
  &\dot\phi' - i\ddot\phi + 2(\beta e^{i\phi} - \alpha e^{-i\phi}) = 0 
 \label{sGmKdV2}\\
  &\dot\phi' + (i/4)\ddddot\phi + (3i/8)\dot\phi^2\ddot\phi - ic\ddot\phi
  + 2(\beta e^{i\phi} - \alpha e^{-i\phi}) = 0,
 \label{sGmKdV3}
\end{align}
\end{subequations}
where a prime denotes a derivative with respect to $v$, , and
$\alpha$, $\beta$, and $c$ are arbitrary constants. Note that all
equations in the hierarchy are invariant to translations in both
coordinates, $u$ and $v$. Setting $\alpha=\beta=i/4$ in
Eq.\ (\ref{sGmKdV1}) yields
\begin{equation}
   \dot\phi' = \sin\phi.
\label{SG}
\end{equation}
This is the sine-Gordon (SG) equation in light-cone coordinates,
$u=(x+t)/2, v=(x-t)/2$, describing the swaying angle $\phi(x,t)$ of a
pendulum at position $x$ and time $t$ along a chain of coupled
pendulums. Setting in Eq.\ (\ref{sGmKdV2}) $\alpha=\beta=-1/4$ and
projecting the equation onto a constant $v$ (\ie considering the case
$\dot{\phi}' = 0$), we obtain the physical-pendulum (PP) equation,
\begin{equation}
  \ddot\phi + \sin\phi = 0,
\label{PP}
\end{equation}
where here $\phi(u)$ describes the swaying angle of a single pendulum
at time $u$. Finally, upon setting $\alpha=\beta=1/16$ and $c=-P/4$,
and projecting onto a constant $v$, Eq.\ (\ref{sGmKdV3}) coincides
with the shape equation for the sheet, Eq.\ (\ref{fold}).

The relation to the sG-mKdV hierarchy enables us to obtain solutions
of Eq.\ (\ref{fold}) from known solutions of the PP equation
(\ref{PP}) or projected solutions of the SG equation (\ref{SG}).
Specifically, we take the ``breather'' solution of Eq.\ (\ref{SG}),
\begin{equation}
  \phi(u,v) = 4\tan^{-1} \left[ \frac{\kappa}{k} \frac{\sin(k(u-v+c_1))}
  {\cosh(\kappa(u+v+c_2))} \right], \ \ \ k^2+\kappa^2=1,
\label{breather}
\end{equation}
describing a standing localized wave in the chain of pendulums.  The
two arbitrary constants, $c_1$ and $c_2$, reflect the two symmetries
of the SG chain under translations in space $x$ and time $t$
(equivalently, along the two light-cone coordinates $u$ and
$v$). Projection of the breather solution onto one of the light-cone
coordinates (\eg $u$ with $v=0$) yields the solution to the
localized-fold profile,
\begin{equation}
  \phi(u) = 4\tan^{-1} \left[ \frac{\kappa}{k} \frac{\sin(k(u+c_1))}
  {\cosh(\kappa(u+c_2))} \right], \ \ \ k^2+\kappa^2=1.
\label{fold_sol}
\end{equation}
Substituting it in Eq.\ (\ref{fold}), we find
\begin{equation}
   k = \half (2+P)^{1/2},\ \ \   \kappa = \half (2-P)^{1/2}.
\label{kkappa}
\end{equation}
The expression for the height profile, $h(u)$, can be obtained in
closed form as well but is postponed until the end of
Sec.\ \ref{sec_K}.

\section{The extra symmetry}
\label{sec_K}

\begin{figure}[t]
\vspace{0.3cm}
\centerline{\resizebox{0.53\textwidth}{!}
{\includegraphics{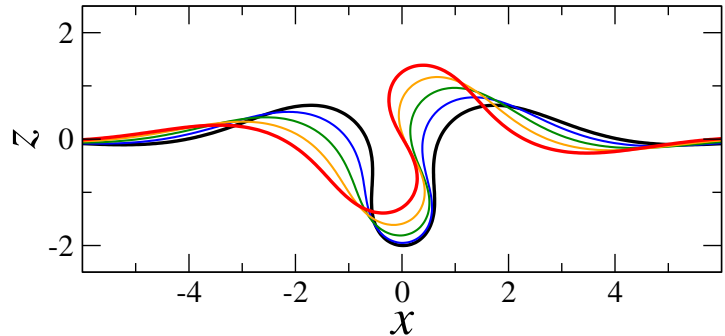}}}
\caption[]{(color online). Shapes of a folded sheet for $P=1$. In
  between the symmetric fold (lowest black curve) and the
  antisymmetric one (uppermost red curve) there is a continuum of
  degenerate shapes having the exact same displacement and energy. The
  shapes were calculated from Eqs.\ (\ref{fold_sol}) and
  (\ref{kkappa}) with $c_2=0$ and $c_1k=0$, $\pi/8$, $\pi/4$,
  $3\pi/8$, and $\pi/2$.}
\label{fig_degenerate}
\end{figure}

The discussion above shows that the folding profiles $\phi(u)$ are
embedded in an autonomous two-dimensional system: the sine-Gordon
system. The obvious translational invariance of the sine-Gordon system
in $v$ entails a corresponding invariance of the folding profiles that
is not obvious: not only can the entire profile be translated along
the sheet, but the oscillatory and decaying parts of the profile can
be shifted independently [cf.\ Eq.\ (\ref{fold_sol})]. Consequently,
in between the symmetric fold ($c_1=c_2=0$) and the antisymmetric one
($c_1k=\pi/2$, $c_2=0$) there exists a continuous family of degenerate
profiles, having the exact same displacement and the exact same
(minimum) energy \cite{aps2012,Rivetti2013}. This is demonstrated in
Fig.\ \ref{fig_degenerate}.

We can directly demonstrate that the specific solution presented in
the preceding section satisfies this symmetry\,---\,\ie that shifting
$v$ in the profile of Eq.\ (\ref{breather}) leaves all energy terms
invariant. The purpose of the current section, however, is to
establish and characterize the extra symmetry at the level of the
original functional to be minimized, Eq.\ (\ref{action}). Thus, any
other solution to the problem will have to satisfy this symmetry as
well.

\subsection{Conservation law}
\label{sec_conserv}

To find the conserved quantity associated with the extra symmetry,
$\cK(\phi,h,p_\phi,p_h)$, we seek the generator of $v$-translations in
the Hamiltonian phase space of Eq.\ (\ref{H}),
\begin{equation}
  \{f,\cK\} = f',
\label{K1}
\end{equation}
where $f(u,v)$ is an arbitrary function expressed in terms of our
canonical variables [Eq.\ (\ref{H})], and $\{\cdots\}$ are Poisson
brackets. In the derivation below we make use of the following: (a)
our knowledge of the first conserved quantity, $\cH$ [Eq.\ (\ref{H})],
and the resulting Hamilton equations (\ref{dynamical}); (b) the set
of equations, analogous to Hamilton's, that derive from $\cK$, \ie
$\{\phi,\cK\}=\pd_{p_\phi}\cK$, $\{h,\cK\}=\pd_{p_h}\cK$,
$\{p_\phi,\cK\}=-\pd_{\phi}\cK$, $\{p_h,\cK\}=-\pd_h\cK$; (c) the fact
that $\phi(u,v)$ should satisfy the SG equation (\ref{SG}). (Note,
however, that we make no use of any particular solution of the SG
equation.) Below we obtain a set of differential constraints on $\cK$,
find a functional form consistent with these, and finally verify that
this $\cK$, (a) generates the desired shift in $v$, and (b) commutes
with $\cH$.

Taking $f=\phi$ in Eq.\ (\ref{K1}) and differentiating both sides with
respect to $u$, we find $(d/du){\{\phi,\cK\}} = \dot\phi' = \sin\phi
= \dot h$. Thus, up to an integration constant, taken as zero,
\begin{equation}
  \pd_{p_\phi}\cK = \phi' = h.
\label{K2}
\end{equation}
We note that the equation $\phi' = h$ provides the connection
between the coordinate $v$ and the actual sheet,
\[
  h dv = \dot\phi du.
\]
It also represents the fact (now in terms of $v$ rather than $u$) that
$\phi$ and $h$ are inter-independent; just as $\dot h=\sin\phi$ along
the $u$ coordinate, $\phi'=h$ along the $v$ coordinate. Applying a
similar procedure with $f=h$ yields $(d/du){\{h,\cK\}} = \dot h' =
(\sin\phi)' = \cos\phi\phi' = h\cos\phi = \dot p_h$. Thus, up to an
integration constant,
\begin{equation}
  \pd_{p_h}\cK = p_h.
\label{K3}
\end{equation}
Next, for $f=p_\phi$, we have $\{p_\phi,\cK\} = p_\phi' =
\dot\phi' = \sin\phi$.  Thus,
\begin{equation}
  \pd_{\phi}\cK = -\sin\phi.
\label{K4}
\end{equation}
Finally, for $f=p_h$, we get $(d/du){\{p_h,\cK\}} = \dot p_h' =
(h\cos\phi)' = \cos\phi h' - h\sin\phi\phi' = p_h\cos\phi
- h^2\sin\phi$. Using Hamilton's equation (\ref{dyn3}), we write this
result as $(d/du)\{p_h,\cK\} = -\dot p_\phi - [(3/2)h^2+P]\sin\phi =
-(d/du)[p_\phi+h^3/2+Ph]$. Thus, up to an integration constant,
\begin{equation}
  \pd_h\cK = p_\phi + \half h^3 + Ph.
\label{K5}
\end{equation}
Gathering the information from Eqs.\ (\ref{K2})--(\ref{K5}), we obtain
an expression for the second conserved quantity,
\begin{equation}
  \cK = \half p_h^2 + h p_\phi + \half h^2 (h^2/4 + P) - (1-\cos\phi).
\label{K}
\end{equation}

We may now verify that this candidate form for $\cK$ satisfies the
requirements (a) and (b) above. By construction, the $\cK$ of
Eq.\ (\ref{K}) satisfies Eqs.\ (\ref{K2})--(\ref{K5}). Thus, it generates
a shift in $v$, fulfilling requirement (a). To address requirement (b)
we may explicitly verify that $\cK$ of Eq.\ (\ref{K}) is a constant of motion,
\begin{equation}
  \{\cK,\cH\} = (\pd_\phi\cK)(\pd_{p_\phi}\cH) - (\pd_{p_\phi}\cK)(\pd_{\phi}\cH)
  + (\pd_h\cK)(\pd_{p_h}\cH) - (\pd_{p_h}\cK)(\pd_{h}\cH) = 0,
\end{equation}
as can be verified by direct substitution using
Eqs.\ (\ref{K2})--(\ref{K5}).  While this $\cK$ bears no obvious
resemblance to $\cH$ of Eq.\ (\ref{H}), the two must nevertheless be
equivalent in generating the fold. The SG equation (\ref{SG}) is
invariant to interchanging $u$ and $v$, and a localized SG solution
along a line of constant $u$ must be equivalent to that along a line
of constant $v$. To verify this duality we form the Hamilton equations
generated by $\cK$,
\begin{subequations}
\label{dynamical_v}
\begin{align}
  \phi' &= \pd_{p_\phi} \cK = h \label{dynv1}\\
  h' &= \pd_{p_h} \cK = p_h \label{dynv2}\\
  p_\phi' &= -\pd_\phi \cK = \sin\phi \label{dynv3}\\
  p_h' &= -\pd_h \cK = -p_\phi - h^3/2 - Ph. \label{dynv4}
\end{align}
\end{subequations}
When the two dynamical systems (\ref{dynamical}) and
(\ref{dynamical_v}) are compared, their equivalence is not
obvious. The replacement $h\leftrightarrow p_\phi$, however, makes
Eqs.\ (\ref{dynv1}) and (\ref{dynv3}) the same as Eqs.\ (\ref{dyn1})
and (\ref{dyn2}), respectively, and Eqs.\ (\ref{dynv2}) and
(\ref{dynv4}) yield Eq.\ (\ref{Euler}) of Euler's {\it elastica}.
Once a single equation for $\phi(v)$ is constructed,
\begin{equation}
  \phi'''' + [(3/2)(\phi')^2 + P]\phi'' + \sin\phi = 0,
\label{fold_v}
\end{equation}
and compared with Eq.\ (\ref{fold}), the equivalence is apparent.

Finally, we use the extra symmetry to obtain closed-form expressions
which could not be derived in earlier works. The height profile,
$h(u)=\int_{-\infty}^u du_1\sin\phi$, is obtained according to
Eq.\ (\ref{dynv1}) by differentiating the breather,
Eq.\ (\ref{breather}), with respect to $v$ and then setting $v=0$. We
get
\begin{eqnarray}
  h(u) &=& -4k\kappa \frac{k\cos\tu_1\cosh\tu_2
  + \kappa\sin\tu_1\sinh\tu_2}
  {k^2\cosh^2\tu_2 + \kappa^2\sin^2\tu_1} \\
  && \tu_1 \equiv k(u+c_1),\ \ \ \tu_2 \equiv \kappa(u+c_2). \nonumber
\end{eqnarray}
Similarly, according to Eq.\ (\ref{dynv2}), we obtain the profile of
displaced fluid mass, $p_h=\int_{-\infty}^u du_1 h\cos\phi$, by
differentiating the breather twice with respect to $v$ and setting
$v=0$. This yields
\begin{eqnarray}
  p_h(u) &=& \frac{2k\kappa}{(k^2\cosh^2\tu_2 + \kappa^2\sin^2\tu_1)^2}
  \left[ 2k\kappa \{ k^2\cosh(2\tu_2) + \kappa^2\cos(2\tu_1) + k^2-\kappa^2 \}
  \cos\tu_1\sinh\tu_2 \right. 
\nonumber\\
  && \left. -\{ (k^2-\kappa^2)[k^2\cosh(2\tu_2) + \kappa^2
  \cos(2\tu_1)] + k^4+6k^2\kappa^2+\kappa^4 \} \sin\tu_1\cosh\tu_2 \right].
\label{ph_sol}
\end{eqnarray}

\subsection{Shape transformation}
\label{sec_shape}

The actual transformation generated by the extra symmetry is
nontrivial.  From Eqs.\ (\ref{dynv1}) and (\ref{dynv2}) we see that a
slight translation $\delta v$ in the extra dimension adds to the angle
of the sheet at each point a small amount proportional to the height
at that point, $\phi(u)\rightarrow\phi(u)+h(u)\delta v$; to the height
it adds a small amount proportional to $p_h$, the fluid mass displaced
up to that point, $h(u)\rightarrow h(u)+p_h(u)\delta v$. In
Fig.\ \ref{fig_degenerate} one can try to follow these small changes
between consecutive curves.

We verify that such an infinitesimal transformation leaves the
mechanical energy of the sheet unchanged. Applying it to the bending
energy, we get
\begin{equation}
  \frac{\delta E_{\rm B}}{\delta v} = \dot\phi\dot h = \sin\phi\dot\phi
  = -(d/du){(\cos\phi)},
\label{inv1}
\end{equation}
which is a total differential, contributing only a boundary term to
the integrated bending energy. That contribution vanishes for our
boundary conditions. For the gravitational energy we get, with the
help of Eq.\ (\ref{dyn4}),
\begin{equation}
  \frac{\delta E_{\rm G}}{\delta v} = h p_h \cos\phi - \half h^3 \sin\phi
  = p_h \dot p_h - \half h^3 \dot h = (d/du){(p_h^2/2 - h^4/8)},
\label{inv2}
\end{equation}
which is a total differential as well. The contribution from the $h^4$
term vanishes for our boundary conditions; the vanishing of the second
boundary contribution is discussed in the next subsection. Thus, both
the bending energy and the substrate energy are {\it separately}
invariant to the transformation. We further confirm that the
functional giving the displacement for a given $P$ is invariant too,
\begin{equation}
  \frac{\delta\Delta}{\delta v} = h\sin\phi = h \dot h = 
  (d/du){(h^2/2)}.
\label{inv3}
\end{equation}
Finally, the variations $\delta\dot h/\delta v=\dot p_h$ and
$\delta\sin\phi/\delta v=h\cos\phi$ are equal thanks to Eq.\
(\ref{dyn4}), ensuring that the transformation does not violate the
geometrical constraint $\dot h=\sin\phi$.

\subsection{Buoyant force}
\label{sec_buoyant}

When defining the system in Sec.\ \ref{sec_system}, we have prescribed
the lateral force per unit length acting on the sheet, $P$, which is
coupled to the lateral displacement, $\Delta$. By contrast, no
explicit term has been introduced to impose an external vertical force
or a certain vertical displacement, apart from the boundary condition
$h(u\rightarrow\pm\infty)=0$. This is equivalent to imposing a zero
vertical force. In the absence of such an external force, the total
vertically displaced fluid mass must vanish. The localized solutions
of the profile equation should automatically satisfy this constraint.

As mentioned above, the mass of vertically displaced fluid up to a
certain point $u$ in the sheet is equal to the $h$-conjugate momentum,
$p_h = \int_{-\infty}^u du_1 h\cos\phi$ [cf.\ Eq.\ (\ref{dyn4})].
Thus, the total displaced fluid mass is $\Mf = p_h(\infty)$.  The fact
that $\Mf=0$ then readily follows from Eq.\ (\ref{dyn3}) applied at
$u\rightarrow\infty$. It is inferred also from the conservation of
$\cK$ when setting $\cK(-\infty)=\cK(\infty)$ in
Eq.\ (\ref{K}). Finally, we explicitly confirm that
$p_h(\pm\infty)\rightarrow 0$ using the closed-form expression for
$p_h$, Eq.\ (\ref{ph_sol}).

\section{Discussion}
\label{sec_discuss}

The physical property of the sheet--fluid composite, underlying the
extra symmetry, is unclear. We have found no physical argument to
explain why the shape transformation described in
Sec.\ \ref{sec_shape} leaves the mechanical energy of the sheet
invariant.  Nonetheless, in the two-dimensional problem, in which we
have embedded our one-dimensional system, the extra symmetry is a
trivial translation invariance along the extra dimension. The
surprising fact that the different energy contributions are {\it
  individually} invariant to the shape transformation generated by
$\cK$ follows as well from an evident symmetry of the sine-Gordon
chain\,---\,the invariance to exchange of its two light-cone
coordinates (which in turn derives from the time-reversal symmetry of
the chain of pendulums). Each energy term is obviously invariant to
translation along the sheet (\ie translation in $u$) and, therefore,
due to the $u\leftrightarrow v$ symmetry, must be individually
invariant also to translation in $v$.

This situation is reminiscent of the phason freedom in quasicrystals
\cite{Levine85,Kalugin85,Lifshitz11}.  In the actual quasicrystal the
degenerate phason states correspond to nontrivial coordinated flips of
atom locations. Yet, once the quasicrystal is represented as a
projection of a higher-dimensional Bravais lattice \cite{Kalugin85},
the phason flips are obtained from simple lattice translations along
the extra dimensions.  An important distinction between the new
symmetry and phasons is that the generated transformation is
continuous rather than discrete. Thus, there are no energy barriers
between the degenerate shapes. Boundary conditions are expected to
remove the degeneracy and will usually favor the symmetric and/or
antisymmetric states \cite{Pocivavsek08}. The corresponding energy
differences, however, should be exponentially small in $\kappa L$.
Therefore, we anticipate important consequences of the continuous
degeneracy for the dynamics of ultra-thin fluid-supported films.

On the face of it, the $\cK$ symmetry arises only from the {\it
  translational} invariance of the two-dimensional host system in
which it is embedded. There is no apparent need for {\it
  integrability} of the host system. Yet, when one attempts to
determine $\cK$ in practice for a non-integrable host system, the
procedure of Eqs.\ (\ref{K1})--(\ref{K5}) fails. The differential
conditions imposed by these equations do not integrate to a
well-defined function of $(\phi,h,p_\phi,p_h)$.  The requirement that
these conditions must define a definite $\cK$ function may then be a
way of formulating the condition for integrability of the host system.

\begin{figure}[tbh]
\vspace{0.3cm}
\centerline{\resizebox{0.7\textwidth}{!}
{\includegraphics{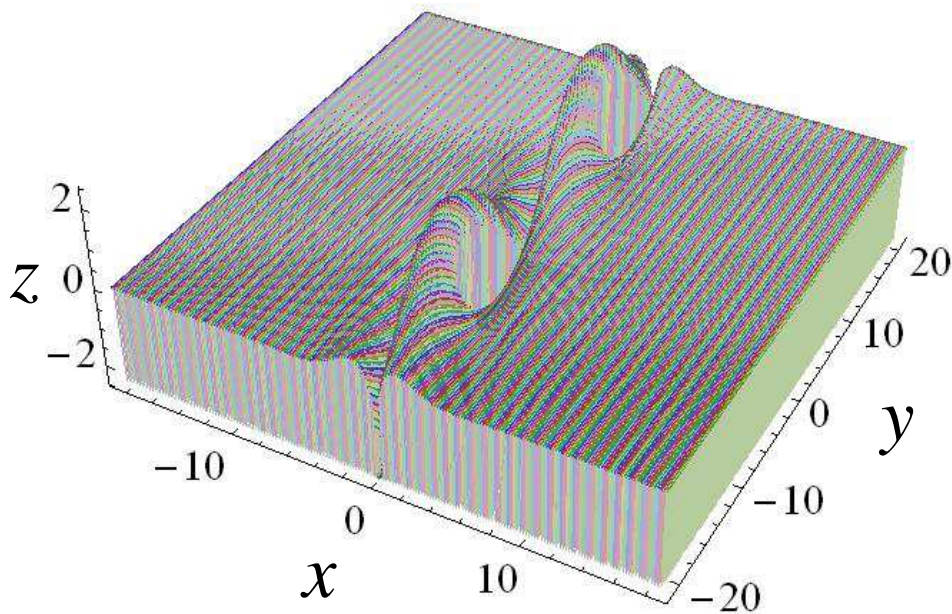}}}
\caption[]{(color online). Shape of a folded sheet whose
  $y$-invariance is broken by a ``soft $v$-mode''. Vertical scale is
  expanded for clarity. Colored stripes are planes of constant
  $u$. Dots are spaced equally in the material, so that the variable
  spacing between them indicates strain. Near edges at $y = \pm 20$
  are symmetric folds pointed downward; two symmetric folds pointed
  upward appear as peaks.  Each $xz$ cut is one of the degenerate
  solutions given by Eq.\ (\ref{fold_sol}) with $c_2=0$ and $c_1$ that
  changes linearly with $y$. The pressure is $P=0.5$.}
\label{fig_twistyfold}
\end{figure}

The floating elastic sheets discussed here occur widely in physical
systems, especially in molecular and nanoparticle monolayers
\cite{Pocivavsek08,Leahy10,Lee08}. Thus we expect the degenerate
shapes described above to have physical implications.  First, one
should in principle see the degeneracy in the model experiments on
macroscopic sheets where this folding was initially demonstrated
\cite{Pocivavsek08}.  However, only the symmetric and antisymmetric
folds were reported.  These sheets had lengths $L$ not much greater
than the width of the deformed region, so that boundary effects not
considered here could have broken the degeneracy. In molecular sheets
\cite{Fischer05} one may imagine a distortion of the $y$-invariant
fold in which the asymmetry parameter varies slowly along the $y$
direction, as demonstrated in Fig.\ \ref{fig_twistyfold}. The extra
energy for such a distortion must vanish as its wavelength goes to
infinity, even though its amplitude is large. Here too, extra effects
may act to select one or another of the asymmetric folds. For example,
self-attraction favors the symmetric fold, since this variant
approaches itself more closely than the others.  On the other hand,
the formation of a fold requires viscous dissipation, enhanced by
shear stress.  This stress is likely greater in the symmetric fold, so
that the antisymmetric fold should be kinetically favored.

The symmetry and degeneracy found here were an indirect and mysterious
consequence of the integrability of the two-dimensional host system.
Though we have discussed it in the context of a particular embedded
system of physical interest, we expect that analogous symmetries occur
elsewhere in the sG-mKdV hierarchy.  These symmetries may ultimately
deepen our understanding of integrability in these systems.  In
physical terms these degenerate deformations and the aforementioned
``soft modes'' associated with them may ultimately shed light on the
experimental puzzles of how molecular folding and unfolding occur in
practice \cite{Lee08,Fischer05,jpc06}.

\begin{acknowledgments}
 We thank Ron Lifshitz for a helpful discussion. This work was
 supported in part by the University of Chicago MRSEC program of the
 NSF under Award Number DMR 0820054.
\end{acknowledgments}


\end{document}